%% file: 00_main.tex
\documentclass[preprint, acmsmall]{acmart}

\AtBeginDocument{%
  \providecommand\BibTeX{{%
    \normalfont B\kern-0.5em{\scshape i\kern-0.25em b}\kern-0.8em\TeX}}}

\usepackage{wrapfig}

\setcopyright{none}
\acmConference[Pre-Print, CSCW 2025]{ACM SIGCHI Conference on Computer-Supported Cooperative Work \& Social Computing 2025}{Bergen, Norway}
    
\acmYear{2025}
\acmDOI{XXXXXXX.XXXXXXX}

\acmSubmissionID{5819}

\newcommand\citetodo[1]{\textcolor{olive}{[CITE]}}

\newcommand{\quoteP}[2][1]{\begin{quote} ``\textit{#2}'' -- P#1 \end{quote}}

\renewcommand{\paragraph}[1]{\vspace{0.2em}\noindent\textbf{\textit{#1}}\hspace*{.3em}}

\usepackage{enumitem}
\usepackage{xspace}
\usepackage{array}
\usepackage{soul}

\begin{document}
\fancyhead[RO]{\fontfamily{LinuxBiolinumT-TLF}\fontsize{8}{10}\selectfont Computer-Supported Cooperative Work \& Social Computing, 2025 --- Author's Preprint}
\fancyhead[LE]{\fontfamily{LinuxBiolinumT-TLF}\fontsize{8}{10}\selectfont Proc. ACM Hum.-Comput. Interact., CSCW 2025}

\title[My Precious Crash Data]{My Precious Crash Data: Barriers and Opportunities in Encouraging Autonomous Driving Companies to Share Safety-Critical Data}

\settopmatter{authorsperrow=4}
\author{Hauke Sandhaus}
\email{hgs52@cornell.edu}
\orcid{0000-0002-4169-0197}
\affiliation{%
  \institution{Cornell University, Cornell Tech}
  \city{New York}
  \state{New York}
  \country{USA}
  \postcode{10044}
}

\author{Angel Hsing-Chi Hwang}
\orcid{0000-0002-0951-7845}
\email{angel.hwang@usc.edu}
\affiliation{%
 \institution{University of Southern California}
  \city{Los Angeles}
  \state{California}
  \country{USA}
  \postcode{90012}
}

\author{Wendy Ju}
\orcid{0000-0002-3119-611X}
\email{wendyju@cornell.edu}
\affiliation{%
  \institution{ Cornell Tech}
  \city{New York}
  \state{New York}
  \country{USA}
  \postcode{10044}
}

\author{Qian Yang}
\orcid{0000-0002-3548-2535}
\email{qianyang@cornell.edu}
\affiliation{%
  \institution{Cornell University}
  \city{Ithaca}
  \state{New York}
  \country{USA}
   \postcode{14853}
 }

\renewcommand{\shortauthors}{Sandhaus et al.}

\begin{abstract}
\input{00_abstract}
\end{abstract}

\ccsdesc[500]{Human-centered computing~Empirical studies in HCI}

\keywords{Autonomous Driving, Data Work, Safety, Industry Practice.}

\settopmatter{printfolios=true} %
\settopmatter{printacmref=false}
\setcopyright{none}
\renewcommand\footnotetextcopyrightpermission[1]{}
\maketitle

\input{01_intro}

\input{02_related_work}

\input{03_method}

\input{04_findings}

\input{05_discussion}
\input{06_conclusion}

\begin{acks}
We express our gratitude to current and former employees at the following automated driving companies and research centers: Argo AI, Bosch, Ford, Intel Mobileye, Mercedes, Motional, Torc Robotics, Tesla, Toyota, Toyota Research Institute, University of Michigan, and Waymo. 
Their willingness to participate in interviews or guide to suitable candidates has been invaluable. We are also grateful to numerous researchers who have offered insights on the early drafts of this work.

This research has been generously supported by the U.S. National Science Foundation under grants IIS-2107111 and IIS-2212431. Additional support for the last author was provided by the Schmidt Futures' AI2050 Early Career Fellowship.
\end{acks}
\bibliographystyle{ACM-Reference-Format}
\bibliography{ref/zotero,ref/manual}

\end{document}

%% file: 00_abstract.tex
Safety-critical data, such as crash and near-crash records, are crucial to improving autonomous vehicle (AV) design and development.
Sharing such data across AV companies, academic researchers, regulators, and the public can help make all AVs safer.
However, AV companies rarely share safety-critical data externally. 
This paper aims to pinpoint why AV companies are reluctant to share safety-critical data, with an eye on how these barriers can inform new approaches to promote sharing.
We interviewed twelve AV company employees who actively work with such data in their day-to-day work.
Findings suggest two key, previously unknown barriers to data sharing: 
(1) Datasets inherently embed salient knowledge that is key to improving AV safety and are resource-intensive.
Therefore, data sharing, even within a company, is fraught with politics.
(2) Interviewees believed AV safety knowledge is private knowledge that brings competitive edges to their companies, rather than public knowledge for social good.
We discuss the implications of these findings for incentivizing and enabling safety-critical AV data sharing, specifically, implications for new approaches to (1) debating and stratifying public and private AV safety knowledge, (2) innovating data tools and data sharing pipelines that enable easier sharing of public AV safety data \textit{and knowledge}; (3) offsetting costs of curating safety-critical data and incentivizing data sharing.

%% file: 01_intro.tex
\section{Introduction}

Sharing data of crashes and near-crashes holds great potential to improve autonomous vehicle (AV) safety research and oversight~\cite{kimEncouragingDataSharing2020, montanaroConnectedAutonomousDriving2019, rawashdehCollaborativeAutomatedDriving2018, rienerCollectiveDataSharing2014}.
Without sufficient safety-critical data, AV safety performance could drop significantly. 
For example, prior research shows that insufficient corner cases in the training data could cause the accuracy of object detection models to drop to 12.8\% mean Average Recall, resulting in unsafe driving conditions~\cite{li_coda_2022}.
This led to the release of CODA\footnote{\url{https://coda-dataset.github.io/}}, a public dataset of real-world road corner cases for object detection in autonomous driving.
Similar movements toward open-sourcing safety-critical data allow researchers across industry and academia to jointly investigate the causes of hazardous AV driving conditions.
The increased availability of such data can also facilitate AV designers and developers to conceive preventative solutions before life-threatening accidents occur.

To date, AV companies rarely share safety-critical data \textit{externally}. 
While policies mandate sharing specific types of AV safety-critical data, companies rarely share beyond the minimal requirements \cite{de_winter_will_2019, fagan_brief_2023}. Recognizing this problem, grassroots movements have started crowdsourcing information about AV crashes and near-crashes \cite{bachman_TeslaDeathsDigital_2023}. AV safety researchers have started curating and sharing data of simulated crashes \cite{hock_how_2018, gulino_waymax_2023, dosovitskiy_carla_2017}. 
Others have started developing data tools that make data sharing easier ~\cite{yurtseverSurveyAutonomousDriving2020,bogdollAddatasetsMetacollectionData2023,ding_survey_2023}. 
These approaches have been highly valuable and impactful. Yet questions remain: Why haven't AV companies started to share their safety-critical data externally and systematically, given the now available data-sharing tools? What other approaches might get companies to do so? What steps are necessary to transform AV safety data into a public good, akin to how vehicle safety features such as seatbelts transitioned to standard safety features~\cite{bellVolvoGiftWorld2019}?%

This paper aims to pinpoint why AV companies are reluctant to share safety-critical data, with an eye on how these barriers can inform new approaches to promote sharing.
Toward this goal, we interviewed twelve AV company employees who work with safety-critical data for AV design and deployment in their day-to-day work.
The interviews focused on understanding their current data management and sharing practices, the challenges and concerns for safety-critical data sharing they have encountered, and the ideal data sharing practices they wish for.

Our interviews identified two key, previously unknown barriers to AV data sharing. Both underscore that AV companies' lack of incentive to share data---more so than the pragmatic challenges around how to share it---as a primary reason behind the rarity of data sharing.
First, an AV company's crash and near-crash data inherently embed knowledge about the machine learning (ML) models and infrastructure that the company uses to \textit{improve} AV safety.
Therefore, such datasets are resource- and knowledge-intensive to curate. 
Data sharing, even \textit{within} a company, is political and fraught.
Second, interviewees believed AV safety knowledge is private knowledge that brings competitive edges to their companies. This perspective leads them to view safety knowledge embedded in data as a contested space rather than public knowledge for social good. 

Re-framing the challenges of AV safety data-sharing as a problem of incentives (\emph{why share?}) rather than a problem of tools (\emph{how to share?}) illuminates new approaches to addressing these challenges. We see a need for AV safety-related communities---academics, policymakers, AV companies, the general public, etc.---to debate and stratify public and private AV safety knowledge. For example, what AV crash data must be shared with regulatory agencies or other AV companies so that similar accidents will not occur?
We see an opportunity for researchers and practitioners to shift data-sharing incentives. First, by removing key negative incentives, researchers can develop data tools and sharing pipelines that make it easier to distinguish public and private knowledge embedded in AV datasets. For example, by building shared virtual platforms that encode crash-prone road scenarios, allowing companies to share knowledge/data about safety-critical \textit{situations} without exposing their preparatory ML models that handle these situations. Second, by exercising strategic influence through standardizing AV safety assessment. Third, responsibility for data sharing could be relieved through strategic collaborations with academic institutions as data intermediaries. Finally, by directly informing policymakers on how to craft data-sharing mandates that balance public safety with industry competitiveness. 
These approaches aim to reframe AV safety-critical data sharing as a public good rather than a solely competitive asset.  
This paper discusses these potential new approaches to improving data sharing, drawn directly from our interview findings.

We make two contributions. First, we reveal the fundamental causes that hinder sharing of safety-critical AV data.
This informs more targeted approaches to motivating data-sharing practice.
Insofar, we highlight three plausible means: defining public vs. private data knowledge, redirecting the design goals of data-sharing tools, and executing incentive programs.
Grounded on these actionable proposals, we call for attention and input from the data work community to make cross-industry-academia data-sharing practices in safety-critical domains more commonplace.

%% file: 02_related_work.tex
\section{Related Work}

To date, there is no one agreed-upon, precise definition of ``\textit{AV safety-critical data}'' across existing literature.
Instead, researchers have used the term to broadly refer to data recorded from AV crashes and near-crash events~\cite{arvinSafetyCriticalEvent2021, klugerIdentificationSafetycriticalEvents2016, wangExploringMechanismCrashes2019, scanlonWaymoSimulatedDriving2021}.
These incidents can occur when various instances of unseen objects, circumstances, and behaviors occur, such as drivers violating traffic rules, demonstrating unsafe driving behaviors, driving under extreme weather conditions, novel objects on roads, or less-common, out-of-context behavior by traffic participants~\cite{scanlonWaymoSimulatedDriving2021, chenExploringMechanismCrashes2021, zhuWhatCanWe2022a, bargman_counterfactual_2017}.
In this paper, we use ``\textit{AV safety-critical data}'' to refer to such data as well.

\subsection{Benefits of  Sharing Safety-Critical Data}
\label{sec:promises}
Availability of safety-critical data is key to facilitating AV safety research, collaboration, and oversight~\cite{koopman_challenges_2016, mirnig_insurers_2019, aaron_h_jacoby_nhtsa_2023, li_coda_2022}.
The automotive research community has advocated for AV companies to share and provide access to such data and called for standardizing data formats such that they can be most useful for improving AV design~\cite{ebel_breaking_2023}.

Existing research has shown the promise of such data for AV safety design, e.g., by adversarial generation of safety-critical events in simulation, and showing the improved performance by re-training on them~\cite{wang_advsim_2021, hanselmann_king_2022}. This data can allow an end-to-end approach to improving autonomous driving, building machine learning (ML) models that use driving context data and generate safer AV driving behaviors~\cite{chen_end--end_2023}. %
Moreover, AV researchers and regulators need AV safety-critical data to investigate reasons for crashes (e.g., Uber's 2018 fatal crash analysis by ~\citet{macrae_learning_2022}), assign responsibilities, and devise strategies for preventing similar incidents.

The automotive research communities depend on real field data to further user-centered design for AVs~\cite{ebel_role_2020}, and maintaining scientific integrity necessitates adherence to open data practices~\cite{ebel_breaking_2023}. This paper investigates whether and why sharing these data types remains limited despite these numerous benefits.

\subsection{The Lack of Safety-Critical Data Sharing}
Despite the above-mentioned promises, abundant evidence shows that safety-critical data remains mostly unavailable to the greater AV research and design community~\cite{guo_is_2018, li_coda_2022}.
Several policies in place uphold this status quo.
The current AV testing policies in Europe and the United States demand minimal crash test data sharing. 
But, such data seldom fully capture underlying factors that cause safety-threatening events~\cite{dan_luu_notes_2024, nordhoff_why_2023, bohm_new_2020}. Often, required data types are limited to descriptive statistics and general information, such as the month when an AV crash occurred, the manufacturer involved, and whether there were injuries. %

Furthermore, officially designated means for data collection do not support recording rich forms of data.
Thus, datasets published through government authorities often lack adequate details to inform AV safety design.
For instance, AV companies are required to report safety-critical events through text documents (e.g., DMV OL 316~\cite{mccarthy_autonomous_2021, californiadm_ol_2020}  and the DMV autonomous vehicle incident web form~\cite{california_department_of_motor_vehicles_autonomous_2024}) in the United States. At most, these forms provide information such as crashes per mile without further details about each incident~\cite{guo_maturity_2022}.
New European policies have mandated that all European-sold vehicles with higher-level automated systems include Event Data Recorders, colloquially known as 'Black boxes.' 
However, data collected through these devices do not capture detailed information about the location, trajectory, time, or context of safety-critical events~\cite{bohm_new_2020, european_transport_safety_council_car_2022, european_commission_-_have_your_say_vehicle_2021}.

\begin{figure}
  \begin{center}
    \includegraphics[height=0.55\textwidth]{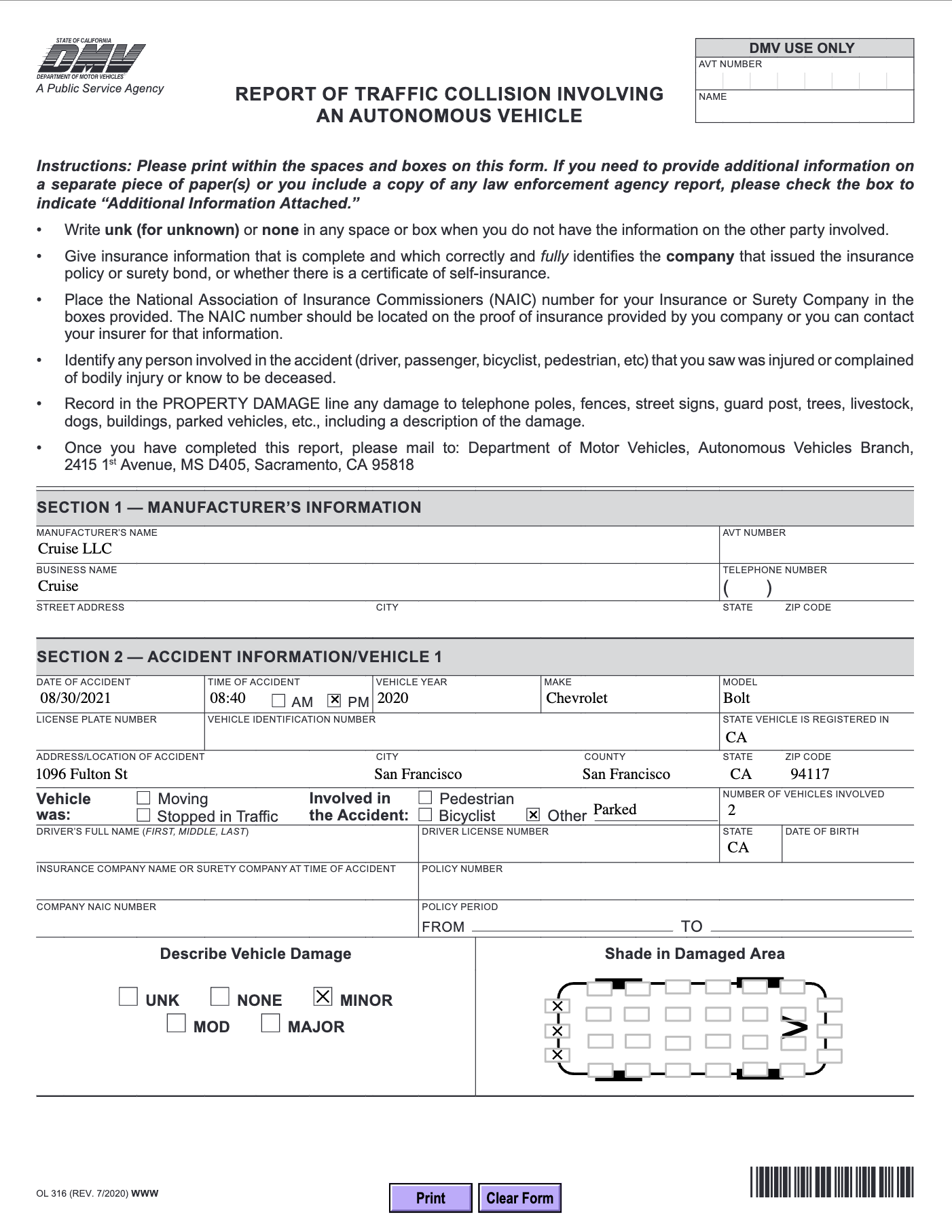}
    \hspace{0.5cm} %
    \includegraphics[height=0.55\textwidth]{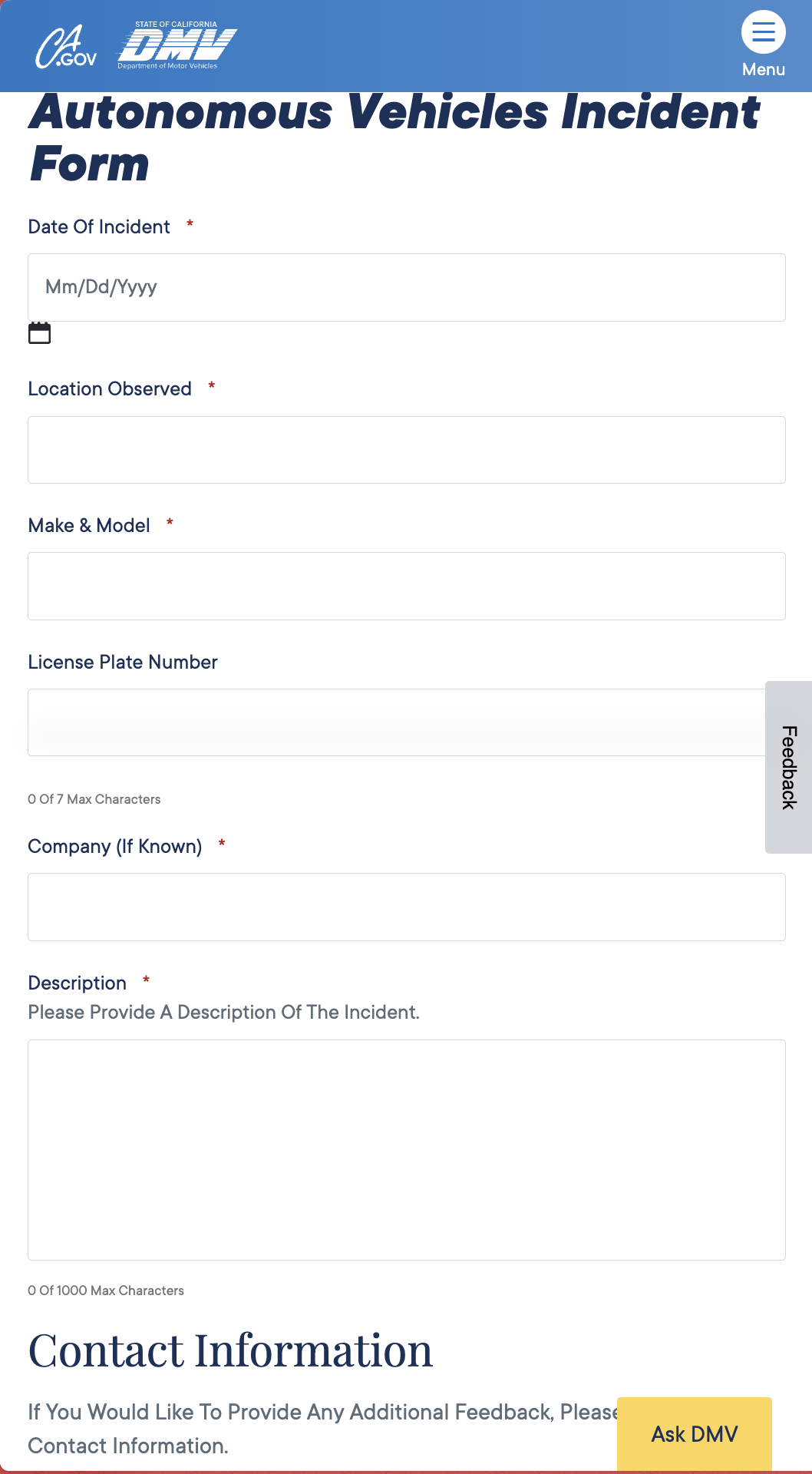}
  \end{center}
  \caption{Left: OL 316, DMV report form for AV collisions (page 1/3). Right: DMV form for the general public to report AV incidents.}
  \Description{OL 316, DMV online report form for autonomous vehicle collisions requires only few basic crash information, such as area of impact and AV developer name. DMV online report form only requires open text description.}
  \label{fig:DMV-form}
\end{figure}

Under these regulations, companies share little beyond the minimal requirements. Recent reviews suggested that datasets shared across the AV industry mostly consist of everyday driving records \textit{only}, rather than those highlighting safety concerns~\cite{yurtseverSurveyAutonomousDriving2020, kimEncouragingDataSharing2020, lee_AreSelfdrivingCars_2023}. 
As a result, even the most comprehensive datasets on AV crashes (see Table 3 from \cite{zhuWhatCanWe2022a}) lack critical details about safety hazards. For example, they do not provide the exact time of the crash, information about the safety driver, speed at the moment of the crash, or micro-location and detailed time-series and movement data. 

Insufficient data-sharing is reflected in two common phenomena: 
First, there are grassroots movements to crowdsource safety-related data. For example, websites are collecting data on deaths involving Tesla's Autopilot \cite{bachman_TeslaDeathsDigital_2023}, and another web tracking automation incidents \cite{daniel_atherton_incident_2023}. 
Second, due to the lack of real safety-critical data, existing studies have resorted to using simulated data to enhance AV safety design. Past research has created crash scenarios by introducing virtual anomalies on roads \cite{bogdoll_perception_2023}, \cite{wang_two_2023}, such as the StreetHazard dataset \cite{hendrycks_scaling_2022}, while some researchers have relied on police reports to roughly reconstructing crash scenarios~\cite{gambiAutomaticallyReconstructingCar2019, gambiGeneratingEffectiveTest2019}.

Together, several studies and reviews of datasets have concluded that the availability of safety-critical data is inadequate for ensuring reliable AV design~\cite{guo_is_2018,li_coda_2022,yurtseverSurveyAutonomousDriving2020, kimEncouragingDataSharing2020, lee_AreSelfdrivingCars_2023}. An overview describing the categories of public and open-sourced safety-critical data sources available to AV researchers is provided in \autoref{tab:available_safety_critical_AV_data}.

\subsection{Barriers to Sharing Safety-Critical Data}\label{ssec:need_understand_barriers}
To encourage AV companies to share data beyond the mandated minimum, we need further understanding of data-sharing obstacles in three aspects.
First, it is essential to explore why the known benefits of data-sharing are not sufficiently motivating. Historically, there are instances where data and knowledge sharing have created collective benefits for the autonomous industry. For example, Volvo's Nils Bohlin relinquished the original safety belt patent, allowing it to be shared freely \cite{bellVolvoGiftWorld2019}. This act of sharing has reportedly saved millions of lives. 

Second, while there are known barriers to general data-sharing in the automotive industry, it remains unclear whether these barriers generalize to sharing AV safety-critical data. The primary barriers to general data sharing fall into several categories. Organizational barriers include lack of expertise or resources needed for effective data sharing \cite{dremelBarriersAdoptionBig2017, panahifar_identifying_2022}, as well as organizational cultures that resist open data practices. Technical barriers encompass challenges in data storage, processing, and movement between systems \cite{rubinfeldAccessBarriersBig2016}, particularly given the scale and complexity of automotive data. Legal and regulatory barriers include privacy legislation, intellectual property concerns, and compliance requirements that limit what data can be shared and how \cite{scheuerman_human_2023, huang_ExaminingModernData_2021, rubinfeldAccessBarriersBig2016}. 
Finally, there might be unknown obstacles that discourage data-sharing practices. We set off our study to reveal and contrast such additional challenges.
Without a fundamental understanding of these data-sharing barriers, it remains challenging to motivate data-sharing practices effectively. Although various solutions have been proposed, their effectiveness is often limited by a lack of comprehensive understanding of the specific obstacles and resistance within the industry.

\subsection{Facilitating Safety-Critical Data Sharing}

\input{tables/open_safety_critical_AV_data}

Emerging work in autonomous driving has focused primarily on developing technical tools to enable sharing of critical safety data. For instance, researchers have introduced federated learning models for sharing AV sensor data \cite{li_privacy-preserved_2022}, blockchain protocols for secure dashcam video exchange \cite{kimEncouragingDataSharing2020}, and other approaches to address scale and privacy barriers in AV data \cite{luckow_automotive_2015, garrido_exploring_2021}. Due to the scarcity of real-world safety data, researchers have also developed methods to generate synthetic safety-critical scenarios by introducing virtual anomalies on roads \cite{bogdoll_perception_2023, wang_two_2023, hendrycks_scaling_2022}, reconstructing crash scenarios from official reports \cite{gambiAutomaticallyReconstructingCar2019, gambiGeneratingEffectiveTest2019}, and enhancing crash datasets with additional contextual data \cite{sinha_crash_2021, zheng_avoid_2023}. While these technical approaches address important aspects of data sharing, they often overlook the organizational and governance processes necessary for successful implementation.

In contrast, other safety-critical industries have developed comprehensive frameworks that integrate both technical and organizational dimensions of data sharing. The aviation industry has evolved sophisticated incident reporting systems that emphasize non-punitive reporting, standardized formats, and clear governance structures \cite{mahajan_CriticalIncidentReporting_2010, walkerRedefiningIncidentsLearn2017}. Their progression from black boxes to Flight Data Monitoring demonstrates how continuous learning can be institutionalized across competing entities through appropriate processes. Similarly, healthcare has established clinical data registries and adverse event reporting structures that carefully balance privacy concerns with the need for collective learning \cite{nebekerEthicalRegulatoryChallenges2017}. CSCW researchers studying work practices across domains have emphasized that successful data sharing depends on human expertise and tacit knowledge \cite{mullerHowDataScience2019, thakkarWhenMachineLearning2022}, often facilitated by trusted intermediaries who manage boundaries between stakeholders \cite{doveOpenDataIntermediaries2023, craginScientificDataCollections2006}. These cross-industry examples suggest that technical solutions alone are insufficient without corresponding organizational frameworks and social practices.

This gap between technical capabilities and organizational readiness has prompted increasing calls for more comprehensive approaches to AV data sharing. Hansen and Pang \cite{hansenOpeningIndustryData2023} note that industrial data sharing requires context-specific approaches rather than direct application of models from academic, governmental, or other sectors. Recent workshops explicitly addressing challenges of data sharing in automotive contexts \cite{ebel_breaking_2023} and methods for obtaining industry data for research \cite{jhaverGettingDataCSCW2023} underscore the growing recognition of this issue. As the types of safety-critical data currently available from public sources show reliance on creative workarounds (summarized in \autoref{tab:available_safety_critical_AV_data}), significant questions remain about how to design effective data-sharing frameworks that respect competitive concerns while maximizing collective safety benefits. These ongoing challenges highlight the need to directly investigate the perspectives of AV industry stakeholders through empirical research, which we address in our interview study.

%% file: tables/open_safety_critical_AV_data.tex
\begin{table*}
\footnotesize
\centering
\renewcommand{\arraystretch}{1.5}
\begin{tabular}{>{\raggedright\arraybackslash}p{1.5cm} 
                >{\raggedright\arraybackslash}p{3.5cm} 
                >{\raggedright\arraybackslash}p{3.5cm} 
                >{\raggedright\arraybackslash}p{3.5cm}}
 \toprule 
 & \textbf{Proactive Identification} & \textbf{Reactive Discovery} & \textbf{Data Enrichment} \\
 \hline
Reports about autonomous vehicles & In the future AV event data recorder triggered events, however data will only be released by court order ~\cite{bohm_new_2020} & AV crash reports~\cite{mccarthy_autonomous_2021, wangExploringMechanismCrashes2019}, police and news reports~\cite{gambiAutomaticallyReconstructingCar2019, gambiGeneratingEffectiveTest2019, holland_verification_2020}, crowd-sourced~\cite{bachman_TeslaDeathsDigital_2023} & Augmenting reports with additional data (e.g. location information~\cite{sinha_crash_2021, zheng_avoid_2023}) \\
 \hline 
Driven vehicle (dashcam)  & Classification of manual-driven vehicle near crash events~\cite{rocky_review_2024, seacrist2018analysis, seacrist2020near, hankey2016description, arvinSafetyCriticalEvent2021} & \emph{Not} possible with monocular dashcam video &Modeling human driver behavior from manual drives~\cite{kandeel_driver_2021, bargman_counterfactual_2017}, sourcing difficult objects from video~\cite{lis_detecting_2019}\\ 
 \hline
Test vehicle & \emph{No} real world autonomous driving safety critical dataset &  \emph{No} real world in-loop autonomous driving safety database & Enriched AV dataset (e.g. extra labels or scenes~\cite{li_coda_2022, blum_fishyscapes_2021, bogdoll_perception_2023}) \\
  \hline
Simulation Environments & Manual scenario design (out of domain objects, in domain objects in unique configurations, weather conditions~\cite{riedmaier_survey_2020, ding_survey_2023, wang_two_2023, hendrycks_scaling_2022}) &  Anomaly in simulation (e.g. crash ~\cite{ding_survey_2023, beck_automated_2023} including adversarial discovery of such~\cite{wang_advsim_2021, ding_learning_2020, hanselmann_king_2022}) & Automated augmentation from ‘real’ road safety critical events  e.g. from dashcams \cite{bashettyDeepCrashTestTurningDashcam2020, scanlonWaymoSimulatedDriving2021, ding_survey_2023}\\ 
\bottomrule
\end{tabular}
\caption{Academic AV researchers rely on sourcing safety-critical driving data from public and open-sourced data.}
\label{tab:available_safety_critical_AV_data}
\end{table*}

%% file: 03_method.tex
\section{Method}

To understand the barriers to sharing safety-critical AV data, we interviewed twelve industry insiders from twelve organizations who actively work on designing, developing, or researching AV safety design with large-scale data. The motivation for our methods is rooted in the rich history of Computer-Supported Cooperative Work research, which has extensively examined the work practices of professionals in software and technology development and increasingly data and machine learning work~\cite{mullerHowDataScience2019,hegerUnderstandingMachineLearning2022,passiTrustDataScience2018,rothschildInterrogatingDataWork2022}.

To this day, very little research has investigated the work practices of autonomous vehicle data workers. We fill this research gap in the present study.
We unfold three primary topics with our interview participants:
(1) how they currently manage and work with AV safety-critical data, and the common challenges they encounter through these current practices; 
(2) what attitudes and rationales they hold toward sharing safety-critical AV data with the greater AV design and research community; 
(3) what more desirable practices of working with data they would like to propose and act toward.

Our interviews took place throughout 2023, a transformative phase in AI marked by innovations such as ChatGPT. Concurrently, the autonomous vehicle sector witnessed volatility, including layoffs at Waymo, the closure of the Argo AI venture, General Motors Cruise reducing their fleet size due to incidents and ongoing delays in Tesla's self-driving package~\cite{bellan_WaymoCuts200_2023, ohnsman_ArgoAIFord_, shakir_TeslaPausesNew_2023}.

\subsection{Participants}

We recruited participants through our extended professional networks to access insiders with substantial insights into the competitive and specialized field of autonomous driving. All participants were from different companies that were committed to developing fully autonomous vehicles.
Our wide range of participants design AVs in either conventional automotive companies or specialized technology companies. 
Participants' demographics and professional experiences were reported in \autoref{tab:participants}.
\input{tables/participants_grouped}

\subsection{Interviews}
We conducted our studies through video interviews, examining participants' current practices and exploring their data needs and barriers. 
The Institutional Review Board of the authors' affiliated institute reviewed and approved the study protocol. The interviews were 50–90 minutes long and were semi-structured. The interview protocol will be available as supplementary material. We ensured the interviewee's anonymity, and the transcripts will remain private.

\subsection{Data Analysis}
We transcribed recordings of the interview sessions for data analysis. Utilizing an iterative inductive coding method~\cite{carpendale_AnalyzingQualitativeData_2017}, two authors extracted initial codes and later used an affinity diagram to organize them into themes. This approach was chosen over grounded theory because it allows themes to emerge from the data without needing a predefined theoretical framework, enabling a more natural identification of patterns relevant to our specific context.

We next applied journey mapping~\cite{endmann_user_2016} to trace the workflows of how these practitioners manage and work with mass-scale AV data. 

We edited quotes within this paper lightly for clarity and readability. We removed speech disfluencies but did not alter the meaning or context of the participants' statements.

%% file: tables/participants_grouped.tex
\newcolumntype{R}[1]{>{\raggedright\arraybackslash}p{#1}}

\begin{table*}[ht]
\centering
\footnotesize
\begin{tabular}{llR{2.5cm}R{2.8cm}R{3.5cm}l}
\toprule
ID  & Gender & Project & Function & Company & Exp. {\tiny (years)} \\
\toprule
\multicolumn{6}{c}{\textbf{Classic Automotive Industry}} \\
\midrule
1  & Male & AV design & Research engineer & Vehicle manufacturer (VM) & $\approx 8$ \\
4* & Male & Distraction modeling & Data-scientist \& researcher & AD department of VM & $\approx 6$ \\
6 & Male & Vehicle perception & Research scientist & AD division of VM & $\approx 7$ \\
8  & Male & Driving performance & Interface researcher & Research division of VM & $ >10$ \\
10 & Male & Behavior modeling & Human factors researcher & Safety research division of VM & $\approx 12$ \\
12 & Male & Vehicle perception & Computer vision engineer & AD department of vehicle supplier &  $\approx 15$ \\ 
\midrule
\multicolumn{6}{c}{\textbf{Specialized Autonomous Driving (AD) Technology}} \\
\midrule
2  & Male & Technical direction & Lead of AD research & AD package provider & $> 10$ \\
3* & Male & Lidar motion estimation & Algorithm engineer & AD software developer & $\approx 6$ \\
5 & Female & Pedestrian behavior & Qualitative researcher & AD software and service developer & $ >10$ \\
7& Male & Cross project alignment & Infrastructure  manager & Mixed-terrain AD provider & $ >15$ \\
9* & Male & AD research & Professor & Driving specialized university & $\approx 8$ \\
11 & Female & AD annotation & Engineering manager & AD vehicle producer & $\approx 5$ \\
\bottomrule
\end{tabular}
\vspace{0.1cm}
\caption{Participant details. Asterisks (*) indicate academic affiliations. Company affiliations are obfuscated.}
\label{tab:participants}
\end{table*}

%% file: 04_findings.tex
\section{Findings}
\label{sec:findings}

Findings from the interviews suggested AV safety-critical data embedded several types of knowledge that were crucial to advancing AV safety design.
As such, our participants were mostly unwilling to make such data publicly available, as they believed it should be private knowledge that yielded a competitive advantage for an AV company.
While AV safety has been a targeted item for fierce competition across the AV industry, this only reinforced practitioners' preferences to keep data as internal resources. 
In contrast to common beliefs in prior literature, which proposed advancing tools to unblock data-sharing barriers, the risks of revealing critical AV safety design knowledge embedded in data are the root causes of practitioners' hesitations.

We structure this Findings section as follows: 
We first provide an overview of our participants' approaches to working with AV safety-critical data, informed by theory of data work practices~\cite{mullerHowDataScience2019}. Next, we examine the barriers widely reported in the literature and weigh them against the experiences of AV data workers to determine if they are 'key' or surmountable (~\autoref{ssec:need_understand_barriers}). This analysis offers essential context as we examine our findings in more detail. 
We then elaborate on the key obstacles to data sharing as identified by our participants, particularly focusing on the risks associated with sharing specific knowledge of AV safety design. Finally, we explore the rationales provided by participants for treating such data as private property, rather than as a public good.

\paragraph{What are the participants' AV safety design work practices?} 
First, most practitioners had constant access to company-owned data sources (see ~\autoref{tab:sourcing_safety_critical_AV_data}).  
While massive data continuously came in as data \textit{streams}, they attained up-to-date data from these company-owned sources on demand, instead of working with one-time, fixed datasets.
All participants predominantly worked with data collected by their own companies. Participant 2 (P2) elaborated, ``Even the biggest companies with a lot of money, vehicles, customers, and users, they still only have their data'' to work with.

\autoref{tab:sourcing_safety_critical_AV_data} categorizes the approaches autonomous driving practitioners use to source safety-critical driving data. We group them into three categories, to easily contrast the proprietary data that practitioners use with the overview of publicly shared data shown in \autoref{tab:available_safety_critical_AV_data}, which academic researchers and policymakers rely on. These approaches are: proactively looking for safety-critical data, e.g., by tuning the vehicle collection parameters remotely to send data in-house for specific scenarios (such as vehicles overtaking in tunnels); discovering safety-critical data reactively, such as annotation from a safety driver (``uncomfortable side swerve''); and enriching existing data to create more safety-critical data, such as generating variations of safety-critical confluences.

All AV data practitioners had access to a basic AV data pipeline, which generally consists of policies for data collection, targeted strategies to capture specific data, methods for organizing and filtering the amassed data, systems for developing and refining AI models, mechanisms to identify and respond to machine learning failures, and strategies for the long-term management and application of the gathered data. Specialized AD companies offer more organizational support for data work, such as implementing robust data governance frameworks, establishing dedicated teams for real-time data monitoring, providing advanced tools for data wrangling and curation, and facilitating ongoing training for AI model development and failure analysis (see \autoref{fig:precious-knowledge-pipeline}). %
Practitioners gradually process data through small, repeated iterative steps involving collecting, organizing, refining, reviewing, oversight, and repurposing data within a well-defined workflow for safety-critical AV data tasks.

\input{tables/sourcing_safety_critical_AV_data}

No practitioner reported using open and external safety-critical data for their AV design. 
Public datasets were of low importance to most interviewees' daily AV design work, except for P3 who worked on autonomous driving algorithm improvements in affiliation with an academic institution. 
Instead, practitioners applied these public datasets for non-design purposes, such as gaining visibility for their work (P3, P7), facilitating training processes (P2, P8, P9), and setting benchmarks for algorithmic performance (P3, P5).

\paragraph{What are surmountable barriers to AV data sharing?}\phantomsection \label{ssec:surmountable_barriers}
Many of the barriers discussed in previous literature are recognized by our AV design insiders but not given the same weight as prior literature; according to them, they are not key in hindering data sharing but false pretexts to hinder sharing data.
Within our interviewees, a lack of motivation and recognition of mutual benefits to data sharing is not in the way; Participants unanimously recognized the benefits of data sharing within their own roles, such as for research and also for the organization, such as through cross-industry collaboration. For example, participants pointed to the economic absurdity of hundreds of companies developing their own test platforms, driving around neighborhoods with unoccupied test vehicles, and collecting data on the same streets repeatedly. Data sharing would allow quicker and more efficient scaling since no AV  company is currently deployed in all areas with a sufficiently sized fleet. \quoteP[6]{It would be a great thing for the research world because you would have data sets in all the different countries and eliminate a lot of waste. Here in the Bay Area, it's crazy how many autonomous cars drive around and just collect the same data over and over. It doesn't really make sense from an overall economical perspective.}
The most commonly discussed data-sharing barrier in street imagery literature is that of privacy concerns; while the interviewees did recognize the innate nature of privacy risks of data sharing, privacy, as protected by law, is not a key barrier. Most participants already protect individual road users’ identities by blurring faces. In fact, they propose even stronger forms of privacy protection or have already deployed privacy-enhancing technologies (PETs) within their infrastructure. With sufficient care, privacy barriers are not insurmountable but maybe just a distraction or \emph{excuse} against data sharing.  Last, while resource constraints on data architecture limit broad access to AV data, existing tools can help overcome these challenges. Companies already implement strategies like data minimization, aggregation, and local computation. These practices, as our participants indicated, could be readily adapted to facilitate resource-efficient data sharing without significant additional overhead. \quoteP[2]{Homomorphic encryption is also - they call it the holy grail of data privacy - it's computationally intensive, but we're working on accelerators to enable these to happen commercially for particular use cases.}

\subsection{Too Entwined to Share: Risks of Publicizing AV Design Knowledge Are Key Barriers to Data-Sharing}
While these challenges are significant, they are not unmanageable; each company has established technical and organizational solutions to mitigate these issues effectively. 
On top of these, practitioners mentioned a more fundamental concern:
\textit{Data inherently carried critical knowledge about how their companies designed and developed AVs.}
The risk of leaking related design knowledge primarily prevented them from sharing data.

Participants pinpointed at least four types of AV design knowledge that were embedded in AV safety-critical data and could be revealed through data sharing; these include: 
(1) how their company defined and operationalized AV safety; 
(2) how they constructed their ML infrastructure; 
(3) where AVs were most susceptible to failure modes;
(4) where a handover of safety liability could take place. 

\begin{figure*}
    \centering
    \includegraphics[width=1\linewidth]{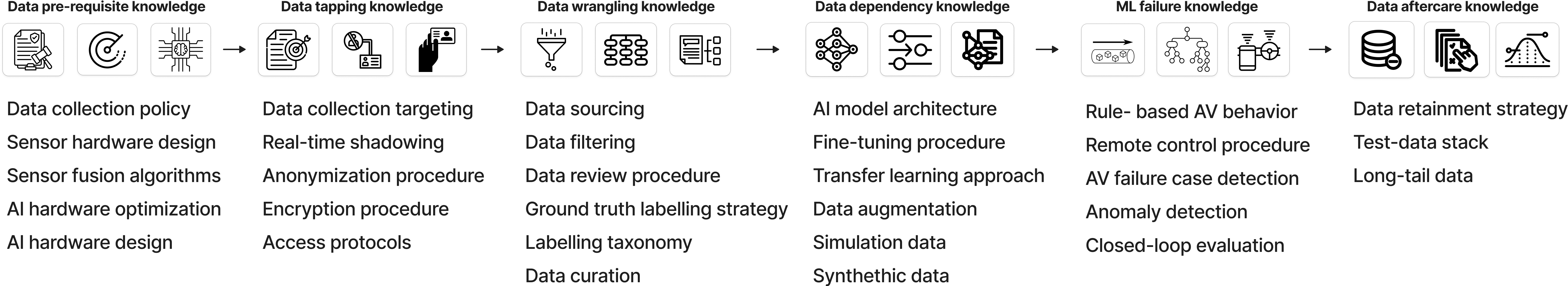}
    \caption{AV developers are reluctant to share data, that encodes precious knowledge, across the industrialized AV data pipeline.}
    \label{fig:precious-knowledge-pipeline}
    \Description{This diagram depicts the stages of knowledge encoded in the AV data pipeline, demonstrating the reluctance of AV developers to share this data. The stages include Data pre-requisite knowledge, Data tapping knowledge, Data wrangling knowledge, Data dependency knowledge, ML failure knowledge, and Data aftercare knowledge. Each stage details various aspects of knowledge such as Data collection policy, Real-time shadowing, Data filtering, AI model architecture, Rule-based AV behavior, and Data retainment strategy, pointing out the comprehensive nature of expertise embedded within the data lifecycle.}
\end{figure*}

These data practices directed changes in the organization of data across all stages. Refer to how data work encodes precious competitive knowledge horizontally across the data pipeline in \autoref{fig:precious-knowledge-pipeline}. 

\paragraph{Stratification of data reveals diverse safety paradigms.}
In the pursuit of safety, AV companies invariably encode substantial knowledge into their data, defining and operationalizing what groups of data constitute safety-critical events. This encoding is not merely a technical process but a reflection of strategic decisions unique to each company's understanding of AV safety.
Participants first highlighted that there was no standard approach toward defining AV safety at scale, %
and thus, how a company defined and operationalized this concept was a strategic decision of its own.
Likewise, there is no consensus on the causes of safety-critical events for autonomous driving.
Participants suggested these ideas were often reflected in how a company sourced data; specifically, when and where data was collected indicated which scenarios were considered safety hazards by an AV design team.
For instance, P12 mentioned a serious AV collision that involved a bicycle or motorcycle rider would likely result in the rider lying on the ground after being hit by the vehicle.
Therefore, identifying a person lying on the ground became a key indicator that helped their team define this specific type of safety-critical event.
However, such contextual knowledge might not be widely shared by all practitioners.

\paragraph{Data sharing risks revealing the machine learning architecture.}
Likewise, participants suggested that how data was collected, stored, and annotated indicated how it would later on be used for ML development.
Oftentimes, this revealed information about key parameters and data structure of ML models. 
Practitioners also feared that ``data carries knowledge downstream''. For example, P12 suggested that ``[one] could get insight on how all sensors are designed based on the data.''
P1 further specified that data could even reveal information about models and prototypes that have not yet been released:

\quoteP[1]{A lot of the data is off internal systems or, like, internal prototypes that haven't ever made it to production or of, you know, internal features, internal software sets, internal sensor sets that haven't been made to production. And so, we do not want any of that information out in public either. So, none of those data sets can be shared.}

Besides, many participants mentioned that their companies created their own suite of tools to pre-process data for their ML models, and AV safety-critical data carried along knowledge about the design of these internal tools.
In P1's company, data was collected and stored in a way that allowed their internal dashboard to readily parse out the time and location where safety-critical events took place.
Alternatively, all data at P11's firm would first go through a system that automatically triaged different types of AV crashes.
Putting together, P8 suggested that AV companies ``built these automated systems because they had consistent goals and consistent perspectives on how they want to use the data'' for their AV design.
On the flip side, attaining data would allow one to infer the metrics, goals, and perspectives held internally at an AV company.

\paragraph{Data sharing risks revealing failure modes.}
According to our participants, AV safety-critical data not only hinted at what constituted their AV design, but it also gave clues to how their design might break and has failed in the past.
Participants mentioned identifying bugs in their ML models was a crucial step in improving the safety performance of AVs.
As such, what their AV safety-critical datasets entailed informed where their models were subject to mal-performance at the moment, how they intended to debug such issues, and eventually, how they advanced AV safety.

\quoteP[11]{But then we need to identify what the issues are. And then a big part of my job was kind of going through all the images that the model failed on. And so if it failed on it, you had to go and triage why did it fail. And if there was a reason for why something failed, you would then go in and add more images into the training set for that specific scenario. And then redo the whole process, and then go over it again to see if the model failed on that scenario again. And if it didn't, then that meant that you fixed that issue. And then you had to go through for the next issue and stuff like that. So a long process of iterating back and forth between evaluation sets.}

\paragraph{Data sharing risks revealing complex liability issues.} 
Finally, participants mentioned that safety-critical data could also entail information about legal liability.
On one hand, owning more data placed more responsibility on AV companies to improve the safety of their vehicles.
On the other hand, ``a lot of these rare events are technically illegal actions by drivers'' (P1).
Many safety-critical incidents took place when drivers were ``not paying attention on the road'' or ``on their phones when they were driving.''
Therefore, participants acknowledged that making such data available would raise complex questions about who should be held accountable for AV safety-critical events.
Consequently, drivers might backfire and become unwilling to grant permission to collect their data.

\label{ssec:Data_not_public_concern}
\subsection{Concerns About Sharing AV Safety-Critical Data as a Public Good}
In addition to the key barrier of potentially sharing sensitive information related to AV design—such as strategic approaches, infrastructure, failure modes, and liability issues—working with safety-critical data transforms it from a readily available asset into a highly valuable resource. 

\paragraph{Data work produces intellectual property that makes data too insightful to share.}
Participants believed their deep engagement with the safety-critical data justified its treatment as private assets.
While they acknowledged on-road safety as a public good, the process of handling data reshaped it as intellectual property and a competitive advantage for AV companies, which caused reluctance for data-sharing.

How do AV companies establish competitive edges through the ways they work with safety-critical data? Our participants indicated three common means:

(1) \textit{Remarking areas in need of new design solutions.} 
Many of our participants admitted pinning down design problems for AV safety improvement could often be more time- and resource-consuming than generating the solutions per se. Knowing what is considered safety-critical is already part of the solution.
As P2 and others elaborated, a variety of contextually specific factors could cause hazardous driving conditions, ranging from light and environmental conditions, unusual objects and animals, and atypical traffic to individual pedestrians' and drivers' behaviors.
Identifying on-point design problems required understanding the full picture of driving scenarios.
The types of data later on used for safety-critical design summarized AV designers' and developers' insights into the primary causes of safety-critical events.

\quoteP[4]{So, this is basically raw data. There's no labeling, for example, for a critical incident. There's no labeling that there was a takeover failure. So, for every analysis that is being made on top of the raw data, you have to define yourself. If you want to know, for example, triggers of a takeover request, then you would need to look into data for these and then you would need to define your labels in that case. There's no annotation that is somehow magically annotating videos.} 

(2) \textit{Identifying safety-critical events from mass data streams}.
Participants suggested that even identifying related safety-critical events \textit{per se} from vast data streams is a resource- and expertise-taxing effort.
As P7 elaborated, there was no handy way to ``look for the top 20 scenarios when a vehicle struggled.''
This is because AV data not only came in with massive volume but also widely differing data structures.
Therefore, having a ``database that is searchable with a fairly wide set of parameters'' would be a tremendous help for AV designers and developers to identify critical events among mass data.

Indeed, most organizations used specialized tools to help with these onerous tasks, such as searching and querying scenarios, labels, and conditions related to safety-critical events.
However, even creating %
``a tool that allows you to query across all those different diverse data and data protocols, considering past and future data, is challenging'' (P7).
Furthermore, P2 suggested the design of each tool embedded ``lessons learned when you encountered a specific AV safety problem.''
Many of our participants' companies had dedicated teams to work on creating effective tools with vast data streams, while most spent a substantial amount of their budgets to procure external tools, hire expert contractors, or even acquire specialized start-ups.
Although several participants indicated their companies held the ultimate goal to ``automate all manual steps in between'' data work (P11), they also admitted the job of identifying safety-critical events will remain highly labor-intensive in the foreseeable future.

(3) \textit{Herding unrevealed AV design insights}.
Participants also acknowledged insights from data did not always become apparent in the first place.
The fact that practitioners took a long time to gauge the value of a dataset was reflected in one of their common work practices -- they seldom absorbed data at once and constantly revisited and wrangled with data.
Before fully figuring out the utility of their data for safety design, most practitioners would hold up with their data given its potential competitive edges or would be advised by their stakeholders (e.g., OEMs) to do so.

\paragraph{Competitive politics in the AV industry drive a secrecy mentality.} 
Participants suggested that fierce competition across the AV industry reinforced the importance of establishing competitive edges through safety-critical data. 
Each believed their company-owned data gave them unique competitive advantages to tackle AV safety and generate one-of-a-kind solutions. While they acknowledged on-road safety as a public good, the process of handling data reshaped it as intellectual property and a competitive advantage for AV companies, which caused reluctance for data-sharing.

\quoteP[2]{The data is the new gold, because using the data, you can develop the solutions. And if you're the first to develop the solutions, it means you're the first to go to market.}

Revisiting the AV companies' AV safety design work practices, all heavily invested in their private data sourcing. All major data work occurs siloed from any publicly available data (contrast ~\autoref{tab:sourcing_safety_critical_AV_data} with~\autoref{tab:available_safety_critical_AV_data}). Participants were entrenched in a mindset that could not envision a future where treating AV safety critical data as public good and competing on different grounds was feasible.

\quoteP[12]{Everybody thinks they have a unique competitive advantage. I guess everybody is hoping they have the only dataset that has all the secret information. Those data are considered valuable. And everybody is not willing to share them.}

Many participants believed their organizations held leading positions in specific types of data work knowledge, although they also acknowledged no company was advantageous in all aspects.
As P2 elaborated, ``each [autonomous] vehicle out there is contributing to solving a particular [AV design] problem.''

Hence, each AV company could claim unique competitive advantages through specific types of data work knowledge, such as: a unique data collection policy (P2, P11), data refinement strategies on hardware and software levels (P2, P4), unique tooling for re-targeting data collection targets in the fleet (P8, P11), leading anonymization and data encryption procedures (P2), data sourcing and filtering tools (P5, P11, P12), leading data labeling procedures and taxonomies (P5), safe rule-based AV algorithms (P3, P7), secret AI architecture (P5, P11), fast closed-loop evaluation methods (P7), transfer learning to scale between countries and cities (P11), field-tested synthetic data augmentation approaches (P7), or knowledge about dealing with the scale of data unlike other industries (P6, P11). In summary, a lot of AV companies felt they were ahead of the competition with some crucial safety-critical data knowledge (See the instances of secretive safety-critical data work practice vertically within the AV data pipeline (\autoref{fig:precious-knowledge-pipeline}).

In conclusion, AV safety has typically been viewed as the frontier of AV design innovation, and being able to act upon safety-critical data needs rapidly has become the key competing ground, as the expensive organizational decisions outlined by one of our interviewees suggested:

\quoteP[11]{Initially they were outsourcing the labeling. But then there was a big push of keeping the data in-house and also investing in an in-house data team. We had our whole called data annotation org, which was 600 700 people, All of the team was in the US, for quality, speed and safety reasons.}

\subsection{Opportunities to Overcome Key Barriers to Data Sharing}

However, even under this competitive landscape, participants still saw the possibilities and advantages of collaborations.
They coined the term ``\textit{untrusted collaborations}'' and believed these types of strategic partnerships would more likely take place when data-sharing was not a prerequisite.
For example, teams at P2's company worked on building federated learning models that allowed different organizations to share lessons learned from data \textit{without} directly sharing their safety-critical data. They elaborated with further details:

\quoteP[2]{
Essentially, it's a way to allow learning from data but in secure enclaves, so that you're not essentially stealing the data or sharing the data. But what you're doing is [...] sharing a common solution, which is maybe a model. And then I'm going to enable that model to be training my data. And then I'm going to benefit from that, that model being improved on my data without you having to share the data. 
}

Participants mentioned similar approaches have been adopted in the healthcare domain, leading to significant, ``30 to 40\% improvements on the performance of these algorithms by learning from these siloed datasets.''  Echoing this potential in the realm of autonomous driving, P6 envisioned a scenario akin to the development of language models like ChatGPT: ``If you could have hundreds of thousands of kilometers in different countries with decent ground truth, you could build something like the ChatGPT, we still don't have that in autonomous driving.''

Only through open data can such expansive and transformative projects be envisioned. According to interviewees, they saw the blame not just on their side, but criticized the academic community for its lack of transparency and pointed to steps toward improved industry-academia data sharing. 

These include increased open access rigor in university research collaborations, with ``centralized data repositories'' (P9), ``explicit open source benchmarks''(P8), ``standardization efforts'' in data sharing (P4), encouragement of ``replicability in research'' (P6) leading to a more general ``appreciation of collective problem solving'' (P2). 
These safety critical data sharing opportunities painted a more plausible future for practitioners to collectively contribute to advancing AV safety design.

%% file: tables/sourcing_safety_critical_AV_data.tex
\begin{table*}
\footnotesize
    \centering
    \renewcommand{\arraystretch}{1.5}
    \setlength{\tabcolsep}{10pt} %
    \begin{tabular}{>{\raggedright\arraybackslash}p{1cm}
                    >{\raggedright\arraybackslash}p{3.3cm}
                    >{\raggedright\arraybackslash}p{3.3cm}
                    >{\raggedright\arraybackslash}p{3.3cm}}
    \toprule
    & \textbf{Proactive Identification} & \textbf{Reactive Discovery} & \textbf{Data Enrichment} \\
    \midrule
    Deployed vehicles & Preset vehicles to capture specific circumstances, collecting data for ongoing interest.\vspace{5pt}\newline Adjust vehicle data collection settings remotely based on current data needs, waiting for specific scenarios to occur. & Document any instances where drivers disengage.\vspace{5pt}\newline Capture bug reports and instances of autonomous vehicle disengagement. & Search through retained data, i.e., filtering in labeled databases, identifying and isolating occasions relevant to ongoing analyses.\\
    \midrule
    Testing vehicles & Conduct controlled drives under safety-critical situations such as varied environmental conditions (hours, weather, locations).\vspace{5pt}\newline Design and execute staged scenarios such as accidents to collect crash data. & Monitor and record any interventions and feedback from supervisors during test drives.\vspace{5pt}\newline Analyze autonomous vehicle-initiated disengagements. & 
    Recreate variations of known safety critical driving situations.\vspace{5pt}\newline
    Manually review test rides to source-related driving behavior. \\
    \midrule
    Simulation environment & Develop simulations from expert insights to model rare but plausible driving scenarios. & Analyze simulations for anomalies (e.g., crashes). & Automatically augment simulations using real road safety-critical events and randomized scenarios (e.g., introduce novel variety, objects). \\
    \bottomrule
    \end{tabular}
    \vspace{5pt}
    \caption{Autonomous driving practitioners source safety-critical driving data using three general approaches, from three sources, in contrast to academic AV researcher (compare with~\autoref{tab:available_safety_critical_AV_data}).}
    \label{tab:sourcing_safety_critical_AV_data}
\end{table*}

%% file: 05_discussion.tex
\section{Discussion}

Findings from the present study show practitioners' reluctance to share safety-critical data is not merely a technical or procedural issue; it is indicative of a deeper need for a paradigm shift. 

Prior work has concentrated on developing tools to facilitate data sharing, yet our findings indicate a different challenge: safety-critical data inherently embeds specific AV safety design knowledge. 
This nature of data causes fundamental obstacles to data sharing, as practitioners fear sharing data would share specific knowledge about AV design knowledge. 
The view of safety-critical data strictly as a competitive advantage rather than public knowledge immobilizes practitioners from sharing data that could significantly advance public AV safety design.

Based on these key findings, we discuss new approaches to motivating data-sharing practices in safety-critical data design. 
We recommend alternative directions for technical solutions, AV safety assessment, and legislative interventions.
Furthermore, we summarize key takeaways for automotive researchers and the data work community and suggest actionable items they can adopt in their research processes going forward.
Together, we envision a future where the competitive landscape of the AV industry can be leveraged for the collective good.

\subsection{Proposed Approaches to Motivating Data-Sharing Practices}
Grounded in our findings and the theory of open data work and open data intermediaries~\cite{mullerHowDataScience2019, doveOpenDataIntermediaries2023}, we suggest targeted strategies to motivate safety-critical data-sharing. These approaches recognize the identified key barriers and opportunities, rather than focusing on the surmountable barriers known in prior literature (see~\autoref{ssec:surmountable_barriers}). 
Specifically, we build on existing efforts (i.e., technical solutions, work practices, and relevant policies) that address this issue and propose new directions for each.
We recommend new approaches to (1) debating and stratifying public and private AV safety knowledge, (2) innovating data tools and data sharing pipelines that enable easier sharing of public AV safety data \textit{and knowledge}; (3) offsetting costs of curating safety-critical data and incentivizing data sharing.

\paragraph{Protecting data-sharing from knowledge-sharing.}
We first suggest that researchers and developers conceive technical solutions that enable sharing data without sharing design knowledge, given that abundant prior work has already been dedicated to developing novel data-sharing tools that address scale and privacy concerns~\cite{luckow_automotive_2015, garrido_exploring_2021}.
Our findings suggest that prior efforts might not directly address practitioners' concerns (i.e., possible leaks of AV design knowledge), as these proposed tools mostly target smoothing the process of data sharing.
This distinction between data artifacts and embedded knowledge is crucial—while sharing raw sensor readings may seem straightforward, our participants revealed that this data implicitly contains proprietary design approaches and decisions. This insight explains why technically sound data-sharing solutions have struggled to gain traction: they address data logistics but not knowledge protection. Developers should now pivot towards methodologies that can decouple safety-critical data from other intricate knowledge it carries, ensuring that data utility is preserved while safeguarding competitive insights, for example, by adopting scenario-based testing~\cite{holland_verification_2020}, for safety-critical instances. %

\paragraph{Assessing AV safety without accessing data.}
While, on the current path, limited data-sharing will likely continue in the near future, advancing AV safety cannot wait. 
We ask whether there can be alternative approaches to leveraging safety-critical data from AV companies without having them directly share their data.
As existing work has made many attempts to simulate environments and scenarios for AV development and testing \cite{hock_how_2018, gulino_waymax_2023, dosovitskiy_carla_2017,xu_safebench_2022}, 
we encourage practitioners to contribute jointly to building standardized simulated platforms for AV safety assessment. Although AV companies may be resistant to sharing data, they have a self-serving interest in shaping the standards by which they are evaluated by the public, their funding sources, and regulatory bodies. 
This might allow practitioners to apply insights they acquire from AV safety-critical data without sharing data first-hand and will also permit robust steps toward AV-safety certification. Establishing clear, recognized benchmarks for safety performance in AV development could have a transformative impact on fostering industry-wide collaboration and advancing public safety. The rapidly advancing field of open-source large language models, which generally rely on closed data while competing on shared benchmarks, could serve as a model for this initiative.

\paragraph{Academic institutions as data intermediaries.} 
Successful open data practices, as seen in nonindustry contexts~\cite{choiCharacteristicsCollaborationEmerging2017,cohoonNegotiatingOpenScience2021, doveOpenDataIntermediaries2023}, highlight the role of intermediaries in ensuring data integrity and accessibility. 
Building on these models, industry stakeholders can leverage the expertise and openness of academic researchers by forming strategic partnerships that involve limited data sharing for specific safety-critical purposes, rather than releasing complete raw data. In these collaborations, academic researchers can take on the role of intermediaries responsible for ensuring that shared data meets open-access and reproducibility standards. This approach addresses the current ``data drought'' in academic research (see \autoref{tab:available_safety_critical_AV_data}) and gradually guides the entire sector toward the standardization of safety-critical data. Our findings indicate that the beneficiaries of such data collaborations should uphold open-data principles; academic partners must be resolute in refusing any agreements that lack commitment to publishing adequate, replicable datasets. Unlike in healthcare—where widely accepted performance benchmarks exist (e.g., cancer prediction scores)—no equivalent standards currently frame the assessment of safety-critical AV data. Our interviews suggest this limited academic-industry collaboration model offers business value for companies while serving public safety interests—several participants specifically noted they would be more open to sharing with academic partners because it offered reputational benefits while limiting direct competitive exposure.

\paragraph{Incentivizing data-sharing through policy frameworks and regulatory approaches.}
Our findings also show great potential for policy interventions to incentivize data-sharing, enhancing the effectiveness of the above-proposed strategies. 
Toward this goal, we first encourage practitioners to reflect on successes in other technological domains, such as healthcare and cybersecurity, where legislative and policy frameworks have facilitated substantial progress \cite{yangDesigningTechnologyPolicy2023}. We ask whether similar initiatives can be modeled after in the AV industry.
Second, we recommend active comparisons across different legislative frameworks as productive practices.
For example, the European Union has advanced a more unified approach to data governance through the EU Data Act~\cite{European-Commission2024-zz}, which establishes structured pathways for the sharing of data that was previously subject to protection. 
By contrast, in the United States, relevant legislation is primarily enacted at the state level. This allows for regulatory flexibility tailored to local contexts, but it also poses challenges for implementing consistent, large-scale data-sharing frameworks across jurisdictions~\cite{fagan_brief_2023,sandhausPrototypingDriverlessVehicle2023}.

Our interviewees consistently highlighted that a lack of incentives remains a major hurdle for data sharing. %
We propose that researchers in data work and transportation should focus on helping policymakers design tiered data-sharing strategies that distinguish between safety-essential knowledge (which should be shared) and competitive design insights (which may remain proprietary). Specifically, researchers can begin by advising policymakers on redesigning the means that guide the collection of minimally required public data (\autoref{fig:DMV-form}) and researching autonomous vehicle safety databases, platforms, and tools. Implementing incentive programs that can help offset the costs of data collection will likely provide immediate motivation for data-sharing.

\subsection{Limitations and Future Work}

Although we aimed for a diverse sample of industry practitioners in autonomous vehicle design, this interview study could be skewed by the interviewees we were able to engage. While all participants were recruited from different companies, we relied on personal networks for referrals, as open sampling proved difficult due to companies' interest in protecting intellectual property. This might bias the findings to a group of participants who are more open to discussing proprietary practices. These interviews also present a Western viewpoint and exclude processes and attitudes in the East and Global South. 
Lastly, the study relied on interviews and could benefit from ethnographic observation of data work practices to cross-verifying the accuracy of the practices and perceptions shared by the participants. 

The present study identified novel key barriers and opportunities to AV data sharing; future work should build from the discussion, explore what types of data can be shared and under which conditions, and specify what safety-critical data needs to be shared for sufficient safety research and oversight, to realign power imbalances. 

%% file: 06_conclusion.tex
\section{Conclusion}
Our study reveals significant barriers that prevent autonomous vehicle (AV) companies from sharing safety-critical data, despite the clear benefits of such sharing for advancing AV safety. Through interviews with industry insiders, we identified two primary obstacles: the inherent embedding of critical knowledge within the data and the perception of safety knowledge as a competitive asset rather than a public good. 

These findings highlight the need for a paradigm shift in how data sharing is approached. Rather than focusing solely on technical solutions to facilitate data exchange, it is essential to address the underlying incentives and strategic concerns of AV companies. We propose concerted efforts from academics, policymakers, and industry practitioners to create technical solutions, policy interventions, and collaborative frameworks to mitigate these barriers.